\documentstyle[opex]{article}

\begin{document}
\title{Fractional Talbot effect in phase space: 
A~compact summation formula}

\author{Konrad Banaszek$^\ast$ and Krzysztof W{\'o}dkiewicz}
\address{Instytut Fizyki Teoretycznej, Uniwersytet Warszawski,
Ho\.{z}a 69, PL--00--681~Warszawa, Poland}
\author{Wolfgang P. Schleich}
\address{Abteilung f\"{u}r Quantenphysik, Universit\"{a}t Ulm,
D-89069 Ulm, Germany}
\email{$^\ast$E-mail address: Konrad.Banaszek@fuw.edu.pl}
\begin{abstract}
A phase space description of the fractional Talbot effect, occurring in
a one--dimensional Fresnel diffraction from a periodic grating, is
presented.  Using the phase space formalism a compact summation formula
for the Wigner function at rational multiples of the Talbot distance is
derived. The summation formula shows that the fractional Talbot image
in the phase space is generated by a finite sum of spatially displaced
Wigner functions of the source field. 
\end{abstract}
\ocis{(070.6760)  Talbot effect}
\begin{OEReferences}
\item\label{Winthrop}
J.~T.~Winthrop and C.~R.~Worthington, ``Theory of Fresnel images.
I. Plane periodic objects in monochromatic light,'' \josa
{\bf 55}, 373--381 (1965).
\item\label{Bastiaans} M.~J.~Bastiaans, ``The Wigner distribution
function applied to optical signals and systems,'' \oc {\bf 25},
26--30 (1978).
\item\label{Averbukh} I.~Sh.~Averbukh and N.~F.~Perelman,
``Fractional revivals: universality in the long--term evolution
of quantum wave packets beyond the correspondence principle
dynamics,'' \pl {\bf A139}, 449--453 (1989).
\item\label{Guigay} J.~P.~Guigay, ``On Fresnel diffraction by
one-dimensional periodic objects, with application to structure
determination of phase objects,'' Opt.\ Acta {\bf 18}, 677--682
(1971).
\item\label{Berry}
M.~V.~Berry and S.~Klein, ``Integer, fractional and fractal
Talbot effects,'' J.~Mod.\ Opt.\ {\bf 43}, 2139--2164 (1996).
\item\label{Parker} 
J.~Parker and C.~R.~Stroud, Jr., ``Coherence and decay
of Rydberg wave--packets,'' \prl {\bf 56}, 716--719 (1986).
\item\label{Yurke}
B.~Yurke and D.~Stoler, ``Generating quantum--mechanical
superpositions of macroscopically distinguishable states via amplitude
dispersion,'' \prl {\bf 57}, 13--16 (1986); A.~Mecozzi and P.~Tombesi,
``Distinguishable quantum states generated via nonlinear birefrigerence,''
\prl {\bf 58}, 1055--1058 (1987).
\item\label{Tara}
K. Tara, G.~S.~Agarwal, and S.~Chaturvedi, ``Production of
Schr\"{o}dinger macroscopic quantum-superposition states in a Kerr
medium,'' \pra {\bf 47}, 5024--5029 (1993).
\item\label{Born}
M.~Born and W.~Ludwig, ``Zur Quantenmechanik des kr\"{a}ftefreien
Teilchens,'' Z.~Phys.\ {\bf 150}, 106--117 (1958).
\item\label{Stifter}
P.~Stifter, C.~Leichte, W.~P.~Schleich, and J.~Marklof,
``Das Teilchen im Kasten: Strukturen in der Wahrscheinlichkeitsdichte,''
Z.~Naturforsch.\ {\bf 52a}, 377--385 (1997).

\end{OEReferences}

\noindent
It is the purpose of this communication to investigate a one dimensional
Fresnel diffraction from a periodic grating and the corresponding fractional
Talbot effect [\ref{Winthrop}]
using the Wigner phase space distribution function.  

We use the following definition of the Wigner distribution function
[\ref{Bastiaans}] for the field amplitude $E(x)$ along the axis $x$
perpendicular to the propagation direction:
\begin{equation}
W(x,u) = \frac{1}{2\pi} \int \mbox{\rm d}y \; E^{\ast}(x + y/2)
e^{iuy} E (x-y/2),
\end{equation}
where $u$ has the meaning of a spatial frequency. We consider a monochromatic
plane wave characterized by its
electric field $ {\cal E}(x;z)=e^{ikz} E(x;z) $ propagated
paraxially along the $z$-axis. At
$z=0$, where the field starts to propagate, there is an infinite
one--dimensional periodic grating with transmittance  $t(x)= t(x+a)$.
The  electric field amplitude after the passage through the grating can be
expanded into the Fourier series:
\begin{equation}
E(x; 0) = t(x)E_{0}= \sum_{n=-\infty}^{\infty} t_{n} e^{2\pi i n x / a},
\end{equation}
where $a$ is the period of the grating and $E_{0}$ is a constant
amplitude of the incident wave. The Wigner distribution
function of this source field is given by:
\begin{eqnarray}
W(x,u; 0) & = & \sum_{n=-\infty}^{\infty} |t_n|^2 \delta (u-nu_0)
\nonumber \\
& + & \sum_{n \neq n'} t_n t_{n'}^{\ast} \exp
[2\pi i (n-n') x/a]
\delta(u - \frac{n+n'}{2} u_0).
\end{eqnarray}
Two different types of contributions can be distinguished in the above
formula.  The first one is a set of parallel positive density stripes
separated by the spatial frequency $u_0=2\pi/a$, given by the reciprocal grid
spacing. Each of them is
generated by a separate Fourier component of the source field.
Coherence between the Fourier components results in nonpositive
oscillatory terms of the Wigner function, located precisely in the middle
between the  contributing frequencies. The nonpositive interference
terms are a consequence of  the linear superposition principle and the
bilinear character of the Wigner function.

The paraxial propagation of the field through a distance
$z$ is described in
the phase space as the following simple transformation of the
Wigner distribution function [\ref{Bastiaans}]:  
\begin{equation} 
W(x,u; z) = W (x - \frac{\lambda z}{2\pi} u, u; 0).
\end{equation}
This formula applied to the Wigner function of the periodic field
gives the following result:
\begin{eqnarray}
W(x,u; z) & = & \sum_{n=-\infty}^{\infty} |t_n|^2 \delta(u-nu_0)
\nonumber \\
\label{Wxuz}
& + &  \sum_{n\neq n'} t_n t_{n'}^{\ast}
\exp[2\pi i (n-n')x/a - 2\pi i (\theta_n - \theta_{n'})]
\delta(u - \frac{n+n'}{2} u_0).
\end{eqnarray}
In the course of propagation the interference terms acquire
additional phase shifts given by
\begin{equation}
\theta_n = \frac{z}{z_T} n^2,
\end{equation}
where $z_T = a^2/2\lambda$ is the Talbot distance. It is 
straightforward to see that for integer multiples
of the Talbot distance the original Wigner function 
of the input field is reproduced. 

At intermediate distances, the phase shifts $\theta_n$ play a
nontrivial role and the structure of the observed Fresnel images
becomes
more complex. Nevertheless, they exhibit an interesting regular
behavior at rational multiples of the Talbot distance. We will now
discuss this effect in terms of the Wigner distribution function.  Let
us denote $z/z_T = p/q$, where $p$ and $q$ are coprime integers.
The main complication in Eq.~(\ref{Wxuz}) are phase factors $\theta_n$
which depend quadratically on $n$. We will simplify it with the help
of an observation used in the studies of quantum wave packets
dynamics [\ref{Averbukh}]:
the exponent $\exp(-2\pi i \theta_n)$ is periodic in $n$ with
the period $l=q/4$ if $q$ is a multiple of 4 and $l=q$ otherwise.
The quadratic phase factor can therefore be represented as a finite
Fourier sum:
\begin{equation}
\exp(-2\pi i \theta_n) = \sum_{s=0}^{l-1} a_s \exp(-2\pi i sn/l)
\end{equation}
with certain coefficients $a_s$, which have been analyzed in
detail in Ref.~[\ref{Averbukh}].
Substitution of the above expression yields:
\begin{eqnarray}
W(x,u; p z_T/q ) & = & 
\sum_{s,s'=0}^{l-1} a_s a_{s'}^{\ast}
\sum_{n, n'=-\infty}^{\infty} t_n t_{n'}^{\ast} 
\exp\left[\frac{2\pi i n}{a} \left(x-\frac{sa}{l}\right)\right]
\nonumber \\
& &
\label{WxumzTkfinal}
\times
\exp\left[-\frac{2\pi i n'}{a} \left(x-\frac{s'a}{l}\right)\right] 
\delta\left(u - \frac{n+n'}{2} u_0\right).
\end{eqnarray}
The interference terms of the Wigner function can now be interpreted as
generated by pairs of Fourier components of the source field shifted in
the position space by rational fractions $sa/l$, 
where $s=0,1,\ldots,l-1$. A simple
rearrangement of the exponent arguments allows one to represent the sum
over $n$ and $n'$
solely in terms of the source field Wigner function:
\begin{equation}
W(x,u; pz_T/q) = \sum_{s,s'=0}^{l-1} a_s a_{s'}^{\ast}
\exp[iu(s'-s)a/l]
W\left(x-\frac{(s+s')a}{2l},u;0\right).
\end{equation}
This is a compact summation formula for the Wigner function at rational
multiples of the Talbot distance. It shows, quite surprisingly, that
$W(x,u; pz_T/q)$ is simply given by a {\em finite\/} sum of spatially
displaced Wigner functions of the source field, with some phase
factors.

Integration of the derived expression for the Wigner function
over $u$ yields
the known formula for the field intensity distribution in the
observation plane:
\begin{eqnarray}
|E(x; p z_T/q)|^2 & = & \int \mbox{\rm d}u \; W (x,u; pz_T/q) 
\nonumber \\
& = & \left| \sum_{s=0}^{l-1} a_s t(x- sa/l; 0) \right|^2 ,
\end{eqnarray}
which shows that the observed Fresnel image can be represented as 
generated by a finite sum of shifted source field amplitudes
[\ref{Guigay}]. A detailed discussion of fractional Talbot images
can be found in Ref.~[\ref{Berry}].

We have discussed a one dimensional Fresnel diffraction from
a periodic grating using the phase space formalism.  We have seen in
this picture, that the whole propagation of the field is encoded in the
interference terms of the Wigner function, generated by coherence
between Fourier components of the source wave.  The variety of the
Fresnel
images is simply a result of a complicated interplay between phases of 
these interference terms. The summation formula derived in this paper
demonstrates that the fractional Talbot effect can be also understood
as an interference between components shifted in the position space by
fractions of the pattern period.

Let us finally note a close analogy between the Talbot effect and the
dynamics of various nonlinear quantum systems. Dynamical phase factors
with a quadratic dependence on the summation index are a common
feature of many problems. These phase factors can be easily revealed in
the equation
of motion for the quantum Wigner function in the $(x,p)$ phase space:
\begin{equation}
W_{\psi}(x,p;t) = \sum_{n,n'} \psi_{n}^{\ast} \psi_{n'}
\exp[i(E_{n}-E_{n'})t/\hbar] W_{nn'}(x,p;0),
\end{equation}
where $\psi_{n}$ are the amplitudes of the initial wave function
$\psi(x;0)$ projected onto the energy eigenstate $\varphi_{n}(x)$,
and where
\begin{equation}
W_{nn'}(x,p) = \frac{1}{2\pi\hbar}
\int \mbox{\rm d}y \, \varphi_{n}^{\ast}(x+y/2) e^{ipy/\hbar}
\varphi_{n'}(x-y/2)
\end{equation}
are the cross Wigner functions of the eigenstates. For many systems,
their eigenergies $E_n$ are given by a quadratic
polynomial of $n$. This can be either an approximate dependence, as it
is for the Rydberg electron wave packets
[\ref{Parker}], or a strict one, as it happens
for an electromagnetic field mode in the Kerr medium 
[\ref{Yurke}]
and a particle in the infinite square well. In all these cases, the
state of the system turns out to be highly regular for times equal to
rational multiples of the characteristic revival time, at which the
system is again in its initial state. This regularity consists in a
formation of a finite superposition of copies of the initial state
generated by a simple transformation. For the wave packets
dynamics, the components are spatially shifted [\ref{Averbukh}]; for
the Kerr medium, the components are rotated in the phase space around
its origin [\ref{Tara}]. 

In particular, a close
similarity exists between the Talbot effect and the fractional
revivals in the infinite square well.  The dynamics of the latter
system is equivalent to an evolution in a free space of an
infinite periodic sequence of displaced initial wave functions
[\ref{Born},\ref{Stifter}].  Furthermore, the propagation of
electromagnetic fields in the paraxial approximation is governed by the
same wave equation as the quantum time evolution.  Therefore,
Fresnel diffraction of classical electromagnetic waves on periodic
gratings is a strict counterpart of the quantum dynamics of a
confined particle.

\section*{Acknowledgments}
The authors have benefited from 
discussions with J. H. Eberly and from a stimulating talk given in Ulm by B.
Rohwedder. This work has been
partially supported by the Polish KBN grant 2 PO3B 006 11 and the Deutsche
Forschungsgemeinshaft.

\end{document}